\documentclass[10pt, conference, letterpaper]{IEEEtran}
\IEEEoverridecommandlockouts
\usepackage[noend]{algpseudocode}
\usepackage{subcaption}
\usepackage{algorithm}
\usepackage{booktabs}
\usepackage{comment}
\usepackage[draft]{hyperref}
\usepackage{amsmath,amssymb,amsfonts}
\usepackage{graphicx}
\usepackage{booktabs}
\usepackage{mathtools}
\usepackage{textcomp}
\usepackage{xcolor}
\usepackage{float}
\usepackage{bm}
\usepackage{bbold}
\usepackage[normalem]{ulem}
\usepackage[inline]{enumitem}
\usepackage{booktabs}
\def\BibTeX{{\rm B\kern-.05em{\sc i\kern-.025em b}\kern-.08em
    T\kern-.1667em\lower.7ex\hbox{E}\kern-.125emX}}

\usepackage{lipsum}

\begin{document}
\title{Scaling Graph-based Deep Learning models to larger networks}
\makeatletter
\newcommand{\newlineauthors}{%
  \end{@IEEEauthorhalign}\hfill\mbox{}\par
  \mbox{}\hfill\begin{@IEEEauthorhalign}
}
\makeatother
\author{\IEEEauthorblockN{Miquel Ferriol-Galmés\textsuperscript{1}, José Suárez-Varela\textsuperscript{1}, Krzysztof Rusek\textsuperscript{1,2}, Pere Barlet-Ros\textsuperscript{1}, Albert Cabellos-Aparicio\textsuperscript{1}}
\IEEEauthorblockA{\textit{\textsuperscript{1}Barcelona Neural Networking center, Universitat Politècnica de Catalunya, Spain}}
\IEEEauthorblockA{\textsuperscript{2}\textit{
AGH University of Science and
Technology, Krakow, Poland}}
}

\maketitle

\begin{abstract}
Graph Neural Networks (GNN) have shown a strong potential to be integrated into commercial products for network control and management. Early works using GNN have demonstrated an unprecedented capability to learn from different network characteristics that are fundamentally represented as graphs, such as the topology, the routing configuration, or the traffic that flows along a series of nodes in the network. In contrast to previous solutions based on Machine Learning (ML), GNN enables to produce accurate predictions even in other networks unseen during the training phase. Nowadays, GNN is a hot topic in the Machine Learning field and, as such, we are witnessing great efforts to leverage its potential in many different fields (e.g., chemistry, physics, social networks). In this context, the Graph Neural Networking challenge 2021 brings a practical limitation of existing GNN-based solutions for networking: the lack of generalization to larger networks. This paper approaches the scalability problem by presenting a GNN-based solution that can effectively scale to larger networks including higher link capacities and aggregated traffic on links.
\end{abstract}


\section{Introduction}
Graph Neural Networks (GNN) have produced groundbreaking applications in many fields where data is fundamentally structured as graphs (e.g., chemistry, physics, biology, recommender systems). In the field of computer networks, this new type of neural networks is being rapidly adopted for a wide variety of use cases \cite{GNNPapersCommNets}, particularly for those involving complex inter-dependencies between different network elements (e.g., performance modeling, routing optimization, resource allocation in wireless networks). In the context of network modeling, unlike previous solutions based on Machine Learning (ML), GNN enables to produce accurate predictions even in networks unseen during the training phase. Nowadays, GNN is a hot topic in the ML field and, as such, we are witnessing great efforts to leverage its potential in many different fields where data is fundamentally represented as graphs (e.g., chemistry, physics, social networks). The Graph Neural Networking challenge~\cite{suarez2021graph} is an annual competition that brings fundamental challenges on the application of GNN to networking.

The 2021 edition of the Graph Neural Networking challenge~\cite{suarez2021graph} brings a fundamental limitation of existing GNNs: their lack of generalization capability to larger graphs. In order to achieve production-ready GNN-based solutions, we need models that can be trained in network testbeds of limited size (e.g., at the vendor’s networking lab), and then be directly ready to operate with guarantees in real customer networks, which are often much larger in number of nodes. In this challenge, participants are asked to design GNN-based models that can be trained on small network scenarios (up to 50 nodes), and after that scale successfully to larger networks not seen before, up to 300 nodes.

This paper presents a possible solution for the Graph Neural Networking challenge 2021, which overcomes the scalability limitations of current GNN models applied to networking. The results show that the proposed model is able to accurately generalize from small-scale networks seen during training (up to 50 nodes) to samples of networks considerably larger (up to 300 nodes) unseen in advance by the model.

\section{Problem statement}
The goal of this challenge is to create a scalable Network Digital Twin (i.e., a network model) based on neural networks, which can accurately estimate QoS performance metrics given a network state snapshot. More in detail, solutions must predict the resulting source-destination mean per-packet delay given: $(i)$ a network topology, $(ii)$ a source-destination traffic matrix, and $(iii)$ a routing configuration (see Figure \ref{fig:nn_model}).

\begin{figure}
    \centering
    \includegraphics[width=\columnwidth]{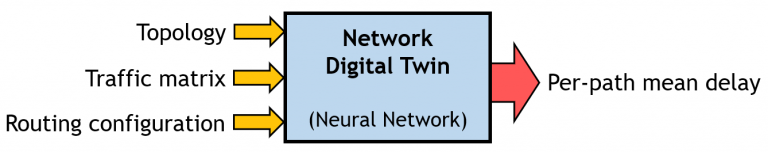}
    \caption{Scheme of the neural network-based solution requested in the Graph Neural Networking challenge~\cite{suarez2021graph}}
    \label{fig:nn_model}
\end{figure}

\section{Background: GNNs} \label{sec:background}
Graphs are used to represent relational information. Particularly, a graph $G \in \{V,E\}$ comprises a set of objects $V$ (i.e., vertices) and some relations between them $E$ (i.e., edges).

GNN~\cite{scarselli2008graph} is a family of NNs especially designed to work with graph-structured data. These models dynamically build their internal NN architecture based on the input graph. For this, they use a modular NN structure that represents explicitly the elements and connections of the graph. As a result, they support graphs of variable size and structure, and their graph processing mechanism is equivariant to node and edge permutation, which eventually endows them with strong generalization capabilities over graphs -- also known as \textit{strong relational inductive bias}~\cite{battaglia2018relational}.

Despite GNN covers a broad family of neural networks with different architectural variants (e.g., \cite{scarselli2008graph, battaglia2016interaction, raposo2017discovering}), most of them share the basic principle of an iterative message-passing phase, where the different elements of the graph exchange information according to their connections, and a final readout phase uses the information encoded in graph elements to produce the final output(s).  We refer the reader to \cite{scarselli2008graph, gilmer2017neural, battaglia2018relational} for a more comprehensive background on GNNs. 

In the context of computer networks, standard GNN architectures, as the one described above, are not applicable, as networks introduce graphs with heterogeneous elements and complex circular dependencies between them. This requires devising a more complex GNN architecture that can adapt to the intricacies of computer networks.

\section{GNN-Based Solution}\label{sec:gnn-model}

This section describes a novel GNN-based solution tailored to accurately model the behavior of real network infrastructures. The model implements a novel three-stage message passing algorithm that explicitly defines some key elements for network modeling (e.g. forwarding devices, queues, paths), and offers support for a wide variety of features introduced in modern networking trends (e.g., complex QoS-aware queuing policies, overlay routing).

\begin{figure}[!t]
    \centering
    \includegraphics[width=\columnwidth]{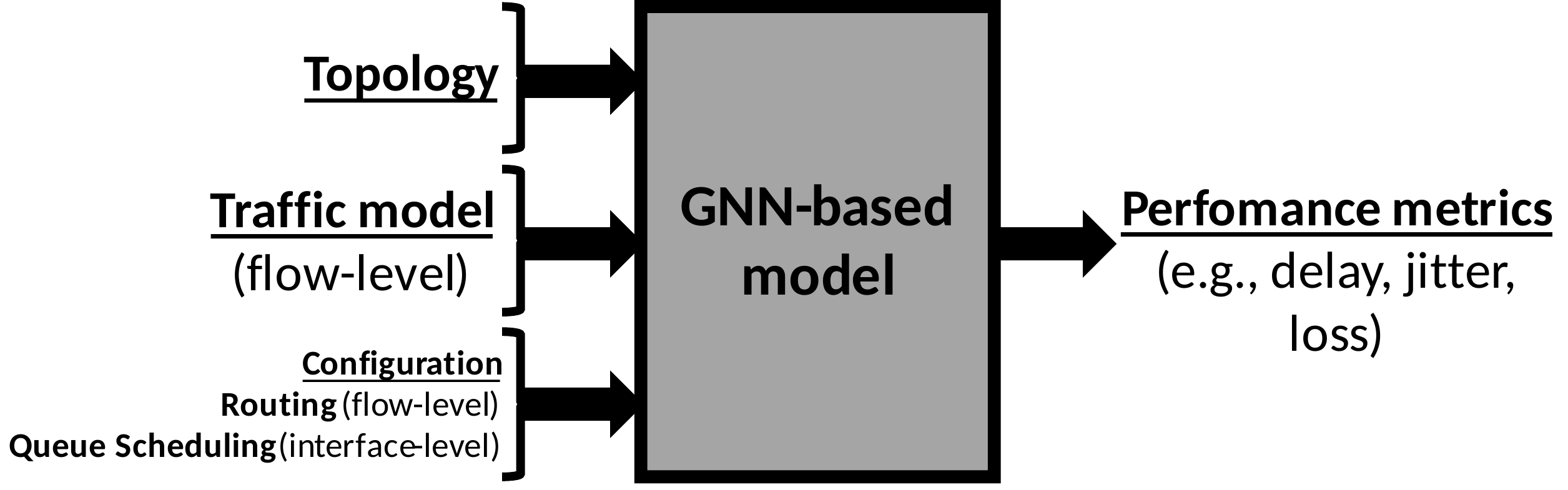}
    \caption{Back-box scheme of the proposed GNN-based model.}
    \label{fig:proposed_gnn_model}
\end{figure}

Figure~\ref{fig:proposed_gnn_model} shows a black-box representation of the proposed GNN-based network model. The input of this model is a network state sample, defined by: a network topology, a set of traffic models (flow-level), a routing scheme (flow-level), and a queuing configuration (interface-level). As output, this model produces estimates of relevant performance metrics at a flow-level granularity (e.g., delay, jitter, losses). Note that, beyond the solution requested in the Graph Neural Networking challenge (Fig.~\ref{fig:nn_model}) the proposed model also supports different queueing policies (e.g., Weighted Fair Queuing, Deficit Round Robin), while in the challenge all forwarding devices implement a FIFO policy.

\subsection{Model Description}\label{sec:model}

The proposed GNN-Based solution pursues two main objectives: $(1)$~Finding a good representation for the network components supported by the model (e.g., traffic intensities, routing), and $(2)$~Exploit scale-independent features of networks, in order to achieve good generalization to larger networks than those seen during training.

\vspace{0.2cm}
\noindent \textit{1) Representing network components and their relationships:}
\vspace{0.1cm}

First, let us define a network as a set of links \mbox{$L = \{l_i: i \in (1,...,n_l)\}$}, a set of queues on $Q = \{q_i: i \in (1,...,n_q)\}$, and a set of source-destination flows $F = \{f_i: i \in (1,...,n_f)\}$. According to the routing configuration, flows follow a source-destination path. Hence, we define flows as sequences with tuples of the queues and links they traverse $f_i=\{(q_{F_q(f_i,0)},l_{F_l(f_i,0)}),...,(q_{F_q(f_i,|f_i|)},$ $l_{F_l(f_i,|f_i|)})\}$, where $F_q(f_i,j)$ and $F_l(f_i,j)$ respectively return the index of the \mbox{$j$-th} queue or link along the path of flow $f_i$. Let us also define $Q_f(q_i)$ as a function that returns all the flows passing through queue $q_i$, and $L_q(l_i)$ as a function that returns the queues injecting traffic into link $l_i$ -- i.e., the queues at the output port to which the link is connected.

\begin{figure}[!t]
\centering
\includegraphics[width=0.9\columnwidth]{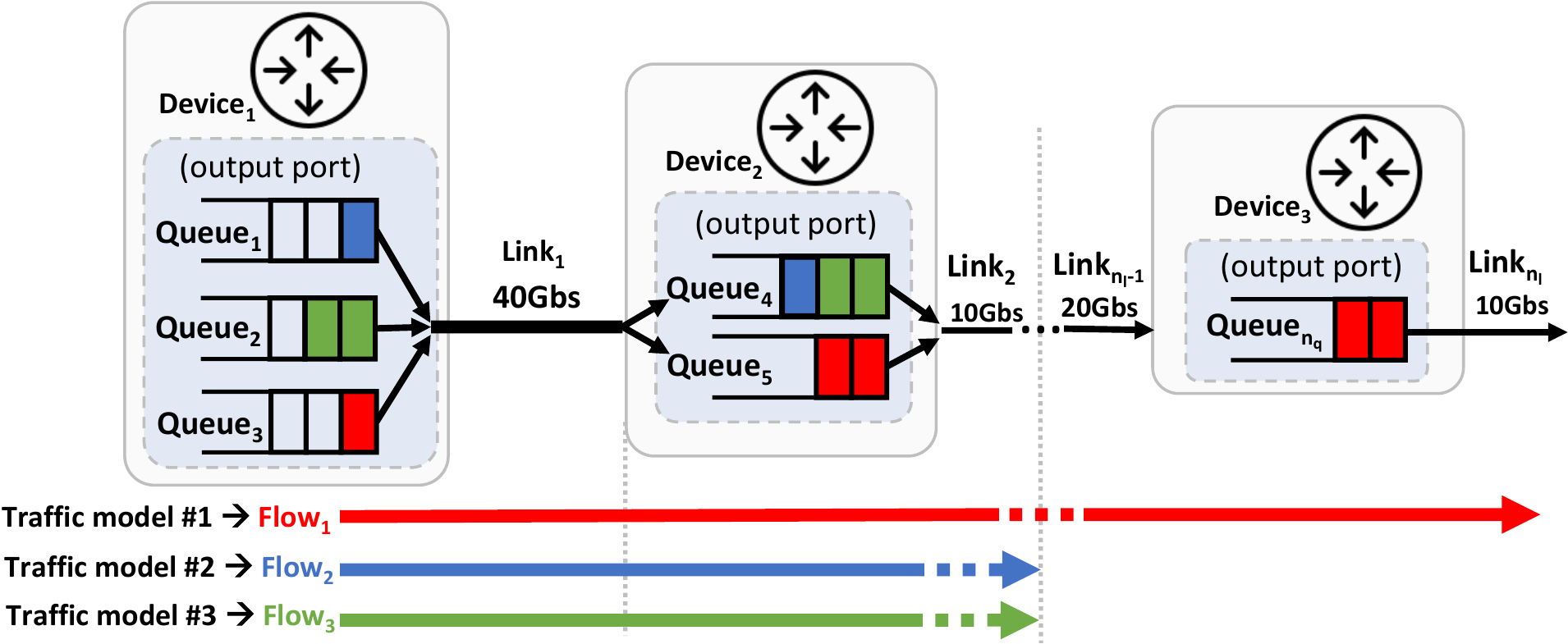}
\caption{Schematic representation of the network model implemented.}
\label{fig:architecture_overview}
\vspace{-0.4cm}
\end{figure}

Following the previous notation, the GNN-based model considers an input graph with three main components: $(i)$ the physical links $L$ that shape the network topology, $(ii)$ the queues $Q$ at each output port of network devices, and $(iii)$ the active flows $F$ in the network, which follow some specific src-dst paths (i.e., sequences of queues and links), and whose traffic is generated from a given traffic model. Figure~\ref{fig:architecture_overview} shows a schematic representation of the network model internally considered, which is derived from the several mechanisms that affect performance in real networks. From this model, we can extract three basic principles:

\begin{enumerate}[label=(\roman*)]
\item The state of flows (e.g., throughput, losses) is affected by the state of the queues and links they traverse (e.g., queue/link utilization).
\item The state of queues (e.g., occupation) depends on the state of the flows passing through them (e.g., traffic model).
\item The state of links (e.g., utilization) depends on the states of the queues at the output port of the link.
\end{enumerate}

Formally, these principles can be formulated as follows:
\begin{align}
h_{f_k} &= g_f(h_{q_{k(0)}},h_{l_{k(0)}},...,h_{q_{k(|f_k|)}},h_{l_{k(|f_k|)}}) \label{eq:g_f}\\
h_{q_i} &= g_q(h_{p_1},...,h_{p_m}), \quad q_i \in p_k, k = 1,...,j \label{eq:g_q}\\
h_{l_j} &= g_l(h_{q_1},...,h_{q_m}), \quad q_m \in L_q(l_j) \label{eq:g_l}
\end{align}

Where $g_f$, $g_q$ and $g_l$ are some unknown functions, and $h_f$, $h_q$ and $h_l$ are latent variables that encode information about the state of flows $F$, queues $Q$, and links $L$ respectively. Note that these principles define a circular dependency between the three network components ($F$, $Q$, and $L$) that must be solved to find latent representations satisfying the equations above.

Based on the previous network modeling principles, we define the architecture of the model (see Algorithm~\ref{alg:architecure}). Our GNN-based model implements a custom three-stage message-passing algorithm that combines the states of flows, queues and links according to Equations~\eqref{eq:g_f}-\eqref{eq:g_l}, thus aiming to resolve the circular dependencies defined in such functions. First, the hidden states $h_l$, $h_q$, and $h_f$ -- represented as n-element vectors -- are initialized with some features (lines \ref{init-l}-\ref{init-p}), denoted respectively by $x_{l_i}$, $x_{q_j}$ and $x_{f_k}$. In our case, we set the initial features of links ($x_l$) as: $(i)$ the link capacity ($C_i$), and \mbox{$(ii)$ the} scheduling policy at the output port of the link (FIFO, SP, WFQ, or DRR~\cite{shreedhar1996efficient}), using one-hot encoding. For the initial features of queues ($x_q$) we include: $(i)$ the buffer size, $(ii)$ the priority level (one-hot encoding), and $(iii)$ the weight (only for WFQ and DRR). Lastly, the initial flow features ($x_f$) are a descriptor of the traffic model used in the flow ($T_i$). Once the states are initialized, the message-passing phase is iteratively executed $T$ times (loop from line~\ref{init-loop}), where $T$ is a configurable parameter. Each message-passing iteration is in turn divided in three stages, that respectively represent the message passing and update of the hidden states of flows $h_f$ (lines~\ref{mp-path-init}-\ref{mp-path-end}), queues $h_q$ (lines~\ref{mp-queue-init}-\ref{mp-queue-end}), and links $h_l$ (lines~\ref{mp-link-init}-\ref{mp-link-end}).

\algnewcommand\algorithmicforeach{\textbf{for each}}
\algdef{S}[FOR]{ForEach}[1]{\algorithmicforeach\ #1\ \algorithmicdo}
\renewcommand{\algorithmicrequire}{\textbf{Input:}}
\renewcommand{\algorithmicensure}{\textbf{Output:}}
\begin{algorithm}[!t] \caption{Internal architecture of the proposed model}
\begin{algorithmic}[1]
\Require{$F$, $Q$, $L$, $x_f$, $x_q$, $x_l$}
\Ensure{$h^T_q$, $h^T_l$, $h^T_f$, $\hat{y_f}$, $\hat{y_q}$, $\hat{y_l}$}
\ForEach {$l \in L$} $h^0_l \gets [x_{l},0...0]$ \EndFor \label{init-l}
\ForEach {$q \in Q$} $h^0_q \gets [x_{q},0...0]$ \EndFor \ForEach {$f \in F$} $h^0_f \gets [x_{f},0...0]$ \EndFor \label{init-p}
\For{t = 0 to T-1} \label{init-loop} \Comment{\footnotesize Message Passing Phase}
    \ForEach {$f \in F$} \label{mp-path-init} \Comment{\footnotesize Message Passing on Flows}
        \ForEach {$(q,l) \in f$}
            \State $h^t_{f} \gets FRNN(h^t_{f},[h^t_q,h^t_l])$ \label{lin:u-flow} \Comment{\footnotesize Flow: Aggr. and Update}
            \State $\widetilde{m}^{t+1}_{f,q} \gets h^t_{f} $ \Comment{\footnotesize Flow: Message Generation}
        \EndFor
    \State $h^{t+1}_{f} \gets h^{t}_{f} $
    \EndFor \label{mp-path-end}
    \ForEach {$q \in Q$} \label{mp-queue-init} \Comment{\footnotesize Message Passing on Queues}
        \State $M_q^{t+1} \gets \sum_{f \in Q_f(q)}  \widetilde{m}^{t+1}_{f,q}$ \Comment{\footnotesize Queue: Aggregation}
        \vspace{0.1cm}
        \State $h^{t+1}_q \gets U_q(h^t_q,M_q^{t+1})$ \Comment{\footnotesize Queue: Update}
        \State $\widetilde{m}^{t+1}_{q} \gets h^{t+1}_{q} $ \Comment{\footnotesize Queue: Message Generation}
    \EndFor \label{mp-queue-end}
    \ForEach {$l \in L$} \label{mp-link-init} \Comment{\footnotesize Message Passing on Links}
        \ForEach {$q \in L_q(l)$}
            \State ${h}^{t}_{l}  \gets LRNN(h^t_l,\widetilde{m}^{t+1}_{q})$  \label{lin:u-link} \Comment{\footnotesize Link: Aggr. and Update}
	    \EndFor
	    \State $h^{t+1}_l \gets {h}^{t}_{l}$
    \EndFor \label{mp-link-end}
\EndFor \label{end-loop}
\State $\hat{y_f} \gets R_f(h^T_f)$ \label{r-path} \Comment{\footnotesize Readout phase}
\State $\hat{y_q} \gets R_q(h^T_q)$ \label{r-queue}
\end{algorithmic}
\label{alg:architecure}
\vspace*{-.1cm}
\end{algorithm}

Finally, functions $R_f$ (line~\ref{r-path}) and $R_q$ (line~\ref{r-queue}) represent independent readout functions that can be respectively applied to the hidden states of flows $h_f$ or queues $h_q$.

\vspace{0.2cm}
\noindent \textit{2) Scaling to larger networks: scale-independent features}
\vspace{0.1cm}

GNNs have shown an unprecedented capability to generalize over graph-structured data~\cite{battaglia2018relational,zhou2018graph}. In the context of generalizing to larger graphs, it is well known that these models keep good generalization capabilities as long as the spectral properties of graphs are similar to those seen during training~\cite{ruiz2020graphon}. In the case of the proposed model, its message-passing algorithm can analogously generalize to graphs with similar structures to those seen during the training phase -- e.g., graphs with a similar number of queues at output ports, or a similar number of flows aggregated in queues. In this vein, generating a representative dataset in small networks, covering a wide range of graph structures, does not imply any practical limitation to then achieve good generalization properties to larger networks. It can be done by simply adding a broad combination of realistic network samples (e.g., with a wide variety of traffic models, routing schemes, queuing policies) -- as in the process described later in section~\ref{subsec:sim-setup}.

However, from a practical standpoint, scaling to larger networks often entails a broader definition beyond the topology size and structure. In particular, there are two main properties we can observe as networks become larger: $(i)$ \textit{higher link capacities} (as there is more aggregated traffic in the core links of the network), and $(ii)$ \textit{longer paths} (as the network diameter becomes larger). This requires devising mechanisms to effectively scale on these two features.

\vspace{0.1cm}
\noindent \textbf{Scaling to larger link capacities:} If we observe the internal architecture of the model (Algorithm~\ref{alg:architecure}), we can find that the link capacity $C$ is only represented as an initial feature of links' hidden states $x_{l_i}$. The fact that $C$ is encoded as a numerical feature in the model introduces inherent limitations to scale to larger capacity values. Indeed, scaling to out-of-distribution numerical values is widely recognized as a generalized limiting factor among all neural networks~\cite{engstrom2019exploring,su2019one}. Thus, our approach is to exploit particularities from the network domain to find scale-independent representations that can define link capacities and how they relate to other link-level features that impact on performance (e.g., the aggregated traffic in the link), as the final goal of this work is to accurately estimate performance metrics (e.g., delay, jitter, losses). Inspired by traditional queuing theory (QT) methods, we aim to encode the relative ratio between the arrival rates on links (based on the traffic aggregated in the link), and the service times (based on the link capacity), thus enabling the possibility to infer the output performance metrics of our model from scale-independent values. As a result, we define link capacities ($Cap_{link}$) as the product of a \textit{virtual reference link capacity} ($C_{ref}$) and a \textit{scale factor} ($S_{f}$):
\begin{equation}
Cap_{link} = C_{ref} * S_{f}
\end{equation}

This representation enables to define arbitrary combinations of scale factors and reference link capacities to define the actual capacity of links in networks. Hence, we introduce the capacity feature ($C_i$) as a 2-element vector defined as $C_i$$=$$[C_{ref}, S_f]$, which is included in the initial feature vector of links ($x_l$). Note that this feature will eventually be encoded in the hidden states of links ($h_l$). In the internal architecture of the GNN-based model (Algorithm~\ref{alg:architecure}), this factor will mainly affect the update functions of flows and links (lines~\ref{lin:u-flow} and \ref{lin:u-link}), as they are the only ones that process directly the hidden states of links ($h_l$). As a result, the RNNs approximating these update functions can potentially learn to make accurate estimates on any combination of $C_{ref}$ and $S_{f}$ as long as these two features are within the range of values observed \textit{independently} for each of them during the training phase (i.e., $S_{f}\in [s_{f_{min}}, s_{f_{max}}]$ and $C_{ref}\in [C_{{ref}_{min}}, C_{{ref}_{max}}]$). Thus, we exploit this property to devise a custom data augmentation method, where we take samples from small networks with limited link capacities and generate different combinations of $C_{ref}$ and $S_{factor}$ that enable us to scale accurately to considerably larger capacities. Note that in this process, the numerical values seen by the model ($C_{ref}$ and $S_{factor}$) are kept in the same ranges
both in the training on small networks and the posterior inference on larger networks, thus overcoming the practical limitation of out-of-distribution predictions~\cite{engstrom2019exploring,su2019one}. More details about the proposed data augmentation process are given in Sec.~\ref{subsec:training}. 

The previous mechanism enables to keep scale-independent features along the message-passing phase of our model (loop lines \ref{init-loop}-\ref{end-loop} in Algorithm~\ref{alg:architecure}), while it is still needed to extend the scale independence to the output layer of the model. Particularly, in this paper, we use the model to predict the flows' delays. Note that the distribution of these parameters can also vary for flows traversing links with higher capacities, thus leading again to out-of-distribution values. Based on the fundamentals of QT, we overcome this potential limitation by inferring delays indirectly from the occupation of queues in the network $O_{q_i}$$\in$$[0,1]$, using the $\hat{y_q}$$=$$R_q(h_q)$ function of Algorithm~\ref{alg:architecure}. Then, we infer the flow delay as a linear combination of the waiting times in queues (inferred from $O_{q_i}$) and the transmission times of the links the flow traverses. Note that a potential advantage with respect to traditional QT models is that the queue occupation estimates produced can be more accurate, especially for complex traffic models resembling real-world traffic.

\vspace{0.1cm}
\noindent \textbf{Scaling to longer paths:} In the internal architecture of the proposed GNN, the path length only affects to the RNN function of line \ref{lin:u-flow} (Algorithm~\ref{alg:architecure}), which collects the state of queues ($h_q$) and links ($h_l$) to update flows' states ($h_f$). The main limitation here is that this RNN can typically see during training shorter link-queue sequences than those it can find then in larger networks, which can potentially have longer paths. As a result, we define $L_{max}$ as a configurable parameter of our model that defines the maximum sequence length supported by this RNN. Then, we split flows exceeding $L_{max}$ into different queue-link sequences that are independently digested by the RNN. To keep the state along the whole flow, in case it is divided into more than one sequence, we initialize the initial state of the RNN with the output resulting from the previous sequence.

\subsection{Simulation Setup} \label{subsec:sim-setup}
To train, validate and test the model we use as ground truth a packet-level network simulator (OMNeT++ v5.5.1~\cite{varga2001discrete}). Each sample is labeled with network performance metrics obtained by the simulator: per-source-destination performance measurements (mean per-packet delay, jitter and loss), and port statistics (e.g., queue utilization, size). To generate these datasets, for each sample we randomly select a combination of input features (traffic model, topology, and queuing configuration).

To test the generalization capabilities to larger networks of our model, we use a wide set of topologies of variable size (from 25 to 300 nodes). All these topologies have been artificially generated using the Power-Law Out-Degree Algorithm described in \cite{palmer2000generating}, where the ranges of the $\alpha$ and $\beta$ parameters have been extrapolated from real-world topologies of the Internet Topology Zoo repository~\cite{6027859}.

In all the previous models, average traffic rates on src-dst flows are carefully set to cover low to quite high congestion levels across different samples, where the most congested samples have $\approx$$3\%$ of packet losses.

\subsection{Training} \label{subsec:training}

We implement the model using TensorFlow. To train the model, we use a custom data augmentation approach that, given a link capacity ($Cap_{link}$), covers a broad combination of $S_f$ and $C_{ref}$ values, in order to eventually make the model generalize over samples with larger link capacities. Particularly, given a link capacity, in some samples, we use low values of $S_f$ with higher values of $C_{ref}$, while in other samples we make it in the opposite way. As an illustrative example, if the model is trained over samples with 1Gbps links, we can represent these capacities in different samples as $Cap_{link}$=$10$*$100Mbps$$=$$1Gbps$, or $Cap_{link}$=$1$*$1Gbps$$=$$1Gbps$. Thus, after training the model should be able to make accurate inferences on samples that combine the maximum $S_f$ and $C_{ref}$ values seen during training -- i.e., $Cap_{link}$=$10$*$1Gbps$$=$$10Gbps$. In practice, this means that the model can be trained with samples with a maximum link capacity of 1 Gbps, and then scale effectively to samples with link capacities up to 10 Gbps. Note that these numbers are just illustrative, while this data augmentation method is sufficiently general to produce in the training dataset wider ranges of $S_f$ and $C_{ref}$ given a maximum link capacity. Thus, it can be potentially exploited to represent combinations leading to arbitrarily larger capacities.

\begin{figure}[!t]
\centering
\includegraphics[width=0.9\columnwidth]{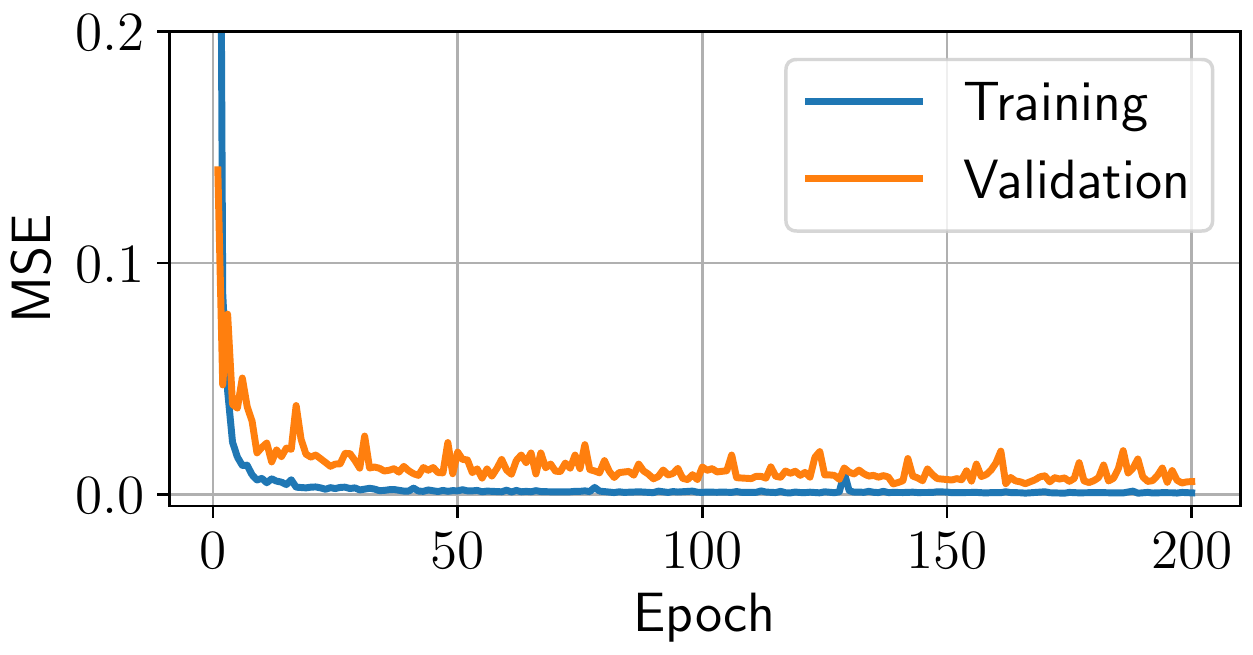}
\caption{Training and evaluation losses over time.}
\label{fig:loss}
\end{figure}

After making some grid search experiments, we set a size of 32 elements for all the hidden state vectors ($h_f$, $h_q$, $h_l$), and $T$=$8$ message-passing iterations. We implement $FRNN$, $LRNN$, and $U_q$ in Algorithm~\ref{alg:architecure} as Gated Recurrent Units (GRU)~\cite{chung2014empirical}, and functions $R_f$ and $R_q$  as 2-layer fully-connected neural networks with ReLU activation functions. Here, it is important to note that the whole neural network architecture of the model (Algorithm~\ref{alg:architecure}) constitutes a fully differentiable function, so it is possible to train the model end-to-end. Hence, all the different functions that shape its internal architecture are jointly optimized during training based on the model's inputs (i.e., the network state samples) and outputs (i.e., the performance metrics).

We train the model for 200 epochs -- with 4,000 samples per epoch -- and set the Mean Squared Error (MSE) as loss function, using an Adam optimizer with an initial learning rate of 0.001. Figure~\ref{fig:loss} shows the evolution of the loss during training on delay estimates (for the training and validation samples), which shows stable learning along the whole training process.

\begin{figure}[!t]
\centering
  \includegraphics[width=0.9\columnwidth]{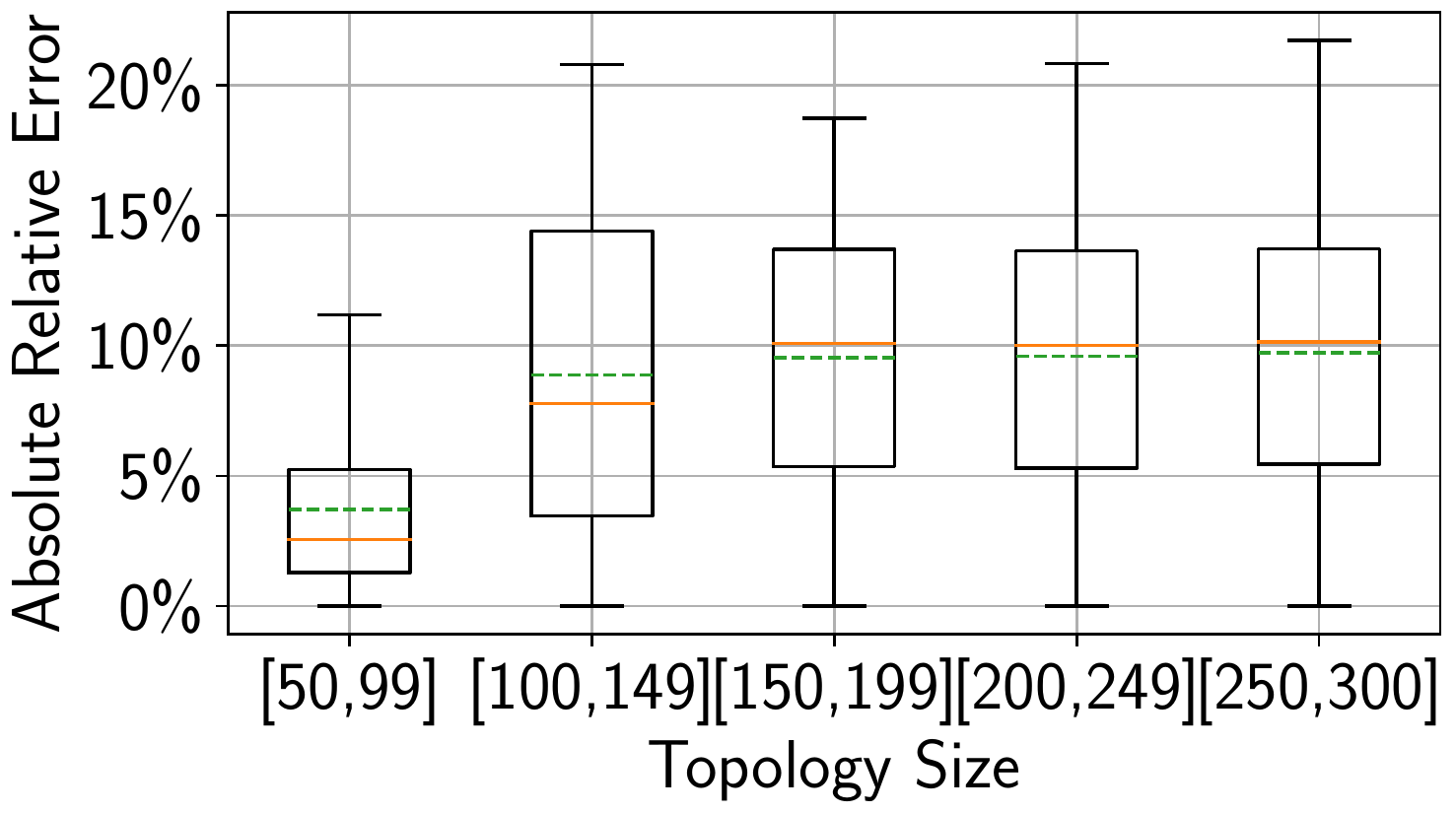}
  \captionof{figure}{Absolute relative error vs. topology size.}
  \label{fig:scalability}
\end{figure}

\section{Evaluation}\label{sec:evaluation}

As we have previously discussed, ML-based network models must generalize to unseen and \emph{larger} networks to become a practical solution. In this vein, the proposed GNN model was carefully designed to address this challenge. Also, note that in the datasets provided for the Graph Neural Networking challenge 2021 all forwarding devices implement a FIFO queue scheduling policy, while the proposed GNN-based model also supports different queueing policies with a variable number of queues and priorities per port. 

\subsection{Generalization to larger networks}
In this section, we evaluate the model in a wide range of networks considerably larger than the ones seen during training. Specifically, the model was trained with topologies between 25 and 50 nodes and tested with topologies from 50 to 300 nodes.

Figure~\ref{fig:scalability} shows how the model generalizes to larger topologies not seen during training. Particularly, the boxplots show the distribution of the absolute relative error with respect to the topology size. As expected, the model obtains better accuracy in topologies that are closer to the ones seen during the training phase (50 to 99 nodes), achieving an average error of $4.5\%$ (green line). As the topology size increases, the average error stabilizes to $\approx$$10\%$.

Generalization is an open challenge in the field of GNN. As we have previously discussed in Sec.~\ref{sec:model}, we have addressed this by using domain-specific knowledge and data augmentation. Particularly, we infer delay/jitter from queues' occupation and apply our scale-independent method to generalize to larger topologies.

\subsection{Inference Speed}

Finally, in this section, we evaluate the inference speed of the proposed GNN-based model. Fast models are especially appealing for network control and management, as they can be deployed in real-time scenarios. For this, we have measured the execution times [Intel(R) Xeon(R) Gold 5220 CPU @ 2.20GHz] in the experiments of the previous section. The results (Table ~\ref{tab:exec_time}) show that the model operate in the order of milliseconds. In particular, it goes from a few milliseconds for small topologies to a few hundred for the larger ones.

\begin{table}[]
\centering
\resizebox{0.8\columnwidth}{!}{%
\begin{tabular}{cccc}
\toprule
           & \multicolumn{3}{c}{Topology Size} \\
           \cmidrule(lr){2-4}
           & [10,30]    & [31,50]     & [51,70]     \\
            \midrule
Exec. Time & 48ms        & 58ms        & 100ms       \\
\bottomrule
\end{tabular}%
}
\caption{Execution  time vs. topology size}
\label{tab:exec_time}
\end{table}
\section{Related Work}

The use of Deep Learning (DL) for network modeling has recently attracted the interest of the networking community. This idea was initially suggested by \emph{Wang, et al.} \cite{wang2017machine}. The authors survey different techniques and discuss data-driven models that can learn real networks. Initial attempts to instantiate this idea use fully-connected neural networks (e.g. \cite{valadarsky2017learning, mestres2018understanding}). Such early attempts do not generalize to networks not seen in training, are not tested with realistic traffic models, and do not model queues. More recent works propose more elaborated neural network models, like Variational Auto-encoders \cite{xiao2018deep} or Convolutional NN \cite{chen2018deep}. However, they have similar limitations. 

Finally, some early pioneering works use GNN in the field of computer networks \cite{geyer2019deeptma, rusek2019unveiling}. However, these models consider a more simplified model of the network,  which do not account for different queuing policies, and the critical property of generalizing to larger networks.

\section{Discussion and Concluding remarks}

In this paper, we have presented a novel GNN-based model that addresses a main limitation of existing ML-based models applied to networks: generalizing accurately to considerably larger networks ($\approx$10x) unseen during training. Particularly, this solution would be suitable for the problem proposed in the Graph Neural Networking challenge 2021~\cite{suarez2021graph}.

\bibliographystyle{IEEEtran}
\bibliography{main}

\begin{thebibliography}{10}
\providecommand{\url}[1]{#1}
\csname url@samestyle\endcsname
\providecommand{\newblock}{\relax}
\providecommand{\bibinfo}[2]{#2}
\providecommand{\BIBentrySTDinterwordspacing}{\spaceskip=0pt\relax}
\providecommand{\BIBentryALTinterwordstretchfactor}{4}
\providecommand{\BIBentryALTinterwordspacing}{\spaceskip=\fontdimen2\font plus
\BIBentryALTinterwordstretchfactor\fontdimen3\font minus
  \fontdimen4\font\relax}
\providecommand{\BIBforeignlanguage}[2]{{%
\expandafter\ifx\csname l@#1\endcsname\relax
\typeout{** WARNING: IEEEtran.bst: No hyphenation pattern has been}%
\typeout{** loaded for the language `#1'. Using the pattern for}%
\typeout{** the default language instead.}%
\else
\language=\csname l@#1\endcsname
\fi
#2}}
\providecommand{\BIBdecl}{\relax}
\BIBdecl

\bibitem{GNNPapersCommNets}
J.~Su{\'a}rez-Varela, M.~Ferriol-Galm{\'e}s, A.~L{\'o}pez, P.~Almasan,
  G.~Bern{\'a}rdez \emph{et~al.}, ``Must-read papers on gnn for communication
  networks,'' \url{https://github.com/BNN-UPC/GNNPapersCommNets}, 2021.

\bibitem{suarez2021graph}
J.~Su{\'a}rez-Varela \emph{et~al.}, ``The graph neural networking challenge: a
  worldwide competition for education in ai/ml for networks,'' \emph{ACM
  SIGCOMM Computer Communication Review}, vol.~51, no.~3, pp. 9--16, 2021.

\bibitem{scarselli2008graph}
F.~Scarselli, M.~Gori \emph{et~al.}, ``The graph neural network model,''
  \emph{IEEE Transactions on Neural Networks}, vol.~20, no.~1, pp. 61--80,
  2008.

\bibitem{battaglia2018relational}
P.~W. Battaglia \emph{et~al.}, ``Relational inductive biases, deep learning,
  and graph networks,'' \emph{arXiv preprint arXiv:1806.01261}, 2018.

\bibitem{battaglia2016interaction}
P.~Battaglia \emph{et~al.}, ``Interaction networks for learning about objects,
  relations and physics,'' in \emph{Advances in neural information processing
  systems}, 2016, pp. 4502--4510.

\bibitem{raposo2017discovering}
D.~Raposo \emph{et~al.}, ``Discovering objects and their relations from
  entangled scene representations,'' \emph{arXiv preprint arXiv:1702.05068},
  2017.

\bibitem{gilmer2017neural}
J.~Gilmer \emph{et~al.}, ``Neural message passing for quantum chemistry,''
  \emph{arXiv preprint arXiv:1704.01212}, 2017.

\bibitem{shreedhar1996efficient}
M.~Shreedhar and G.~Varghese, ``Efficient fair queuing using deficit
  round-robin,'' \emph{IEEE/ACM Transactions on networking}, vol.~4, no.~3, pp.
  375--385, 1996.

\bibitem{zhou2018graph}
J.~Zhou \emph{et~al.}, ``Graph neural networks: A review of methods and
  applications,'' \emph{arXiv preprint arXiv:1812.08434}, 2018.

\bibitem{ruiz2020graphon}
L.~Ruiz, L.~Chamon, and A.~Ribeiro, ``Graph neural networks and the
  transferability of graph neural networks,'' \emph{Advances in Neural
  Information Processing Systems}, vol.~33, 2020.

\bibitem{engstrom2019exploring}
L.~Engstrom, B.~Tran, D.~Tsipras, L.~Schmidt, and A.~Madry, ``Exploring the
  landscape of spatial robustness,'' in \emph{International Conference on
  Machine Learning}, 2019, pp. 1802--1811.

\bibitem{su2019one}
J.~Su, D.~V. Vargas, and K.~Sakurai, ``One pixel attack for fooling deep neural
  networks,'' \emph{IEEE Transactions on Evolutionary Computation}, vol.~23,
  no.~5, pp. 828--841, 2019.

\bibitem{varga2001discrete}
A.~Varga, ``Discrete event simulation system,'' in \emph{European Simulation
  Multiconference (ESM)}, 2001, pp. 1--7.

\bibitem{palmer2000generating}
C.~R. Palmer and J.~G. Steffan, ``Generating network topologies that obey power
  laws,'' in \emph{IEEE Global Telecommunications Conference (GLOBECOM)},
  vol.~1, 2000, pp. 434--438.

\bibitem{6027859}
S.~Knight, H.~Nguyen \emph{et~al.}, ``The internet topology zoo,'' \emph{IEEE
  Journal on Selected Areas in Communications}, vol.~29, no.~9, pp. 1765
  --1775, 2011.

\bibitem{chung2014empirical}
J.~Chung, C.~Gulcehre, K.~Cho, and Y.~Bengio, ``Empirical evaluation of gated
  recurrent neural networks on sequence modeling,'' \emph{arXiv preprint
  arXiv:1412.3555}, 2014.

\bibitem{wang2017machine}
M.~Wang \emph{et~al.}, ``Machine learning for networking: Workflow, advances
  and opportunities,'' \emph{IEEE Network}, vol.~32, no.~2, pp. 92--99, 2017.

\bibitem{valadarsky2017learning}
A.~Valadarsky \emph{et~al.}, ``Learning to route,'' in \emph{ACM workshop on
  hot topics in networks}, 2017, pp. 185--191.

\bibitem{mestres2018understanding}
A.~Mestres \emph{et~al.}, ``Understanding the modeling of computer network
  delays using neural networks,'' in \emph{ACM SIGCOMM BigDAMA workshop}, 2018,
  pp. 46--52.

\bibitem{xiao2018deep}
S.~Xiao \emph{et~al.}, ``Deep-q: Traffic-driven qos inference using deep
  generative network,'' in \emph{ACM SIGCOMM Workshop on Network Meets AI \&
  ML}, 2018, pp. 67--73.

\bibitem{chen2018deep}
X.~Chen \emph{et~al.}, ``Deep-rmsa: A deep-reinforcement-learning routing,
  modulation and spectrum assignment agent for elastic optical networks,'' in
  \emph{2018 Optical Fiber Communications Conference and Exposition (OFC)},
  2018, pp. 1--3.

\bibitem{geyer2019deeptma}
F.~Geyer and S.~Bondorf, ``Deeptma: Predicting effective contention models for
  network calculus using graph neural networks,'' in \emph{IEEE INFOCOM}, 2019,
  pp. 1009--1017.

\bibitem{rusek2019unveiling}
K.~Rusek, J.~Su{\'a}rez-Varela \emph{et~al.}, ``Unveiling the potential of
  graph neural networks for network modeling and optimization in sdn,'' in
  \emph{ACM Symposium on SDN Research}, 2019, pp. 140--151.

\end{thebibliography}

\end{document}